\documentclass[aps,onecolumn]{revtex4}
\usepackage{morefloats}
\usepackage{float}
\usepackage{graphicx}
\usepackage{dcolumn}
\usepackage{amsmath}
\usepackage{subfigure}
\usepackage{epsfig}
\usepackage{times}
\input{epsf}

\newcommand{\bra}[1]{\langle #1|}

\newcommand{\nn}[1]{|#1|^{2}}
\newcommand{\ket}[1]{|#1\rangle}
\newcommand{\braket}[2]{\langle #1|#2\rangle}

\begin{document}


\title{Analysis of a Quantum Nondemolition Measurement Scheme Based on Kerr Nonlinearity in Photonic Crystal Waveguides}


\author{Ilya Fushman}
\altaffiliation[Also at: ]{Department of Applied Physics, Stanford
University, Stanford, CA 94305}
\email[]{ifushman@stanford.edu}
\author{Jelena Vu\v{c}kovi\'{c}}
\altaffiliation[Also at: ]{Department of Electrical Engineering,
Stanford University, Stanford, CA 94305}




\date{\today}

\begin{abstract}

We discuss the feasibility of a quantum nondemolition measurement (QND) of photon number based on cross phase modulation due to the Kerr effect in Photonic Crystal Waveguides (PCW's). In particular, we derive the equations for two modes propagating in PCW's and their coupling by a third order nonlinearity. The reduced group velocity and small cross-sectional area of the PCW lead to an enhancement of the interaction relative to bulk materials.  We show that in principle, such experiments may be feasible with current photonic technologies, although they are limited by material properties. Our analysis of the propagation equations is sufficiently general to be applicable to the study of soliton formation, all-optical switching and can be extended to processes involving other orders of the nonlinearity.

\end{abstract}

\maketitle
\clearpage

In this paper we focus on the feasibility of realizing the QND photon number measurement proposed in \cite{ref:KerrY} in PCW's. QND measurements are important in a variety of quantum information processing techniques as well as quantum state preparation. Although our investigation was motivated by quantum information processing in photonic crystal (PhC) on-chip networks, a detector with the necessary sensitivity may prove to be valuable on its own. We show that the reduction of group velocity and small interaction volumes in PCWs lead to an effective enhancement of the third order nonlinearity and, theoretically, make experiments with high quality structures and attainable laser power levels feasible.\\

	We consider the case of a signal pulse from both a photon number emitter and a coherent state. A typical single photon source is an InGaAs quantum dot (QD) coupled to a PhC cavity as in \cite{ref:Dirk2005}. The radiative lifetime of such QD's coupled to PhC cavities is $\approx 0.2-1 ns$. In this experiment, a weak signal pulse is combined with a strong coherent probe in one arm of a Michaelson interferometer. The probe acquires a phase shift that is directly proportional to the signal photon number, and the signal pulse is retained for further use. The main impediment to this measurement is the small value of the nonlinearity and the relatively large photon absorption in semiconductor materials.\\
	
	The Kerr effect is a third order nonlinearity ($\chi^{(3)}$) and can be described by a weak intensity dependent refractive index ($n_2 I$) as $3\chi^{(3)}=c n^{2} n_2$, where $c$ is the speed of light and $n$ is the refractive index of the material and I is the Electric field intensity \cite{ref:Boyd}. We will focus on Alluminum Gallium Arsenide (AlGaAs), which has a reasonable $\chi^{(3)}$ and $n_2 \approx 1.5 10^{-13} \frac{cm^2}{W}$ at a wavelength of 1500 nm and a high refractive index $n \approx 3.4$ \cite{ref:Aitchison,ref:AlGaAs_Ho}. The large refractive index is attractive for the fabrication of PhC devices and can be combined with our current QD sources. Another material system of interest is Indium Gallium Arsenide Phosphide (InGaAsP), which has $n_2 \approx -5.9 \times 10^{-12} \frac{cm^2}{W} $ at 1545 nm \cite{ref:all_optical_GaInAsP}. An order of magnitude estimate of the phase due to the nonlinearity can be gained by expanding the index as: $\tilde{n} \approx n+n_{2} I$.  The relative dielectric constant to first order in intensity is then $\epsilon \approx n^{2}+2 n n_{2} I$, and the wave vector (k) in the bulk material becomes $k =k_{0}+\Delta k = \frac{\omega}{c} \sqrt{\epsilon}$. The acquired phase over a distance L for a signal and probe photons with wavelength $\lambda_{s},\lambda_{p}$ and $N_{s}$ signal photons, is then  $ \frac{4 \pi^2 c \hbar n_{2}}{\lambda_{s}\lambda_{p}\tau_{s}A} N_{s} $, which for an area of $A = 1 \mu m^2$, length $L=100 \mu m$, a lifetime of $\tau_s = 1 ns$ and $\lambda_{p,s} \approx 1.5 \mu m$ gives $\Delta\Phi \approx 8 \times 10^{-13} \times N_{s}$. The sensitivity of the interferometer is given by its signal to noise ratio (SNR). The noise in interferometry with coherent states comes from the photon partition noise (shot noise) at the input beamsplitters of the interferometer. It can be shown that in the case of a coherent probe and signal in a number state, the SNR is $(4 \phi^{2}_{s} N_p)^{-1}$, where $N_p$ is the probe average photon number, and $\phi_s$ is the phase due to a single signal photon \cite{ref:MQO}. This means that $N_{p} > 10^{23}$ probe photons are needed for the bulk experiment in order to overcome the shot noise when the probe is in a coherent state and the signal is in a number state. This requires a source that can produce a $48 kJ$ pulse with a nanosecond width. \\
	
	PCW's offer two improvements for such a measurement. First, the group velocity of the pulse propagating in a PhC waveguide is reduced to $v \approx 10^{-2} - 10^{-3} \times c$ \cite{ref:JelaSlowLight,ref:IBMSlowLight}, and the local intensity increases by this factor. The effective propagation length of the waveguide increases as well, although the losses associated with material absorption and scattering should also increase with the longer effective length. This enhancement merely allows us to make smaller structures. Pulse contraction also means that longer probe pulses can be generated by the external pulse, and will shrink to the desired width. Secondly, the area of the PhC waveguide is of order $\left(\frac{\lambda}{n}\right)^2$. A combination of these effects leads to an overall enhancement of $\left(\frac{c}{v_g}\right)^2\frac{A_{bulk}}{A_{PhC}} \approx  2\times 10^5$, for a PhC area of $(250 nm)^2$ and $\frac{v}{c}=100$. This reduces the energy requirement to $\approx 10^{-6} J$ in a nanosecond pulse. This figure does not take into account material absorption and waveguide scattering losses, which are generally the main obstacle to such a measurement. We show that with material absorption parameters found in literature and the probe at wavelengths above the half-bandgap of the semiconductors, the experiments could be attempted in a PhC device. Our analysis of the propagation equations is sufficiently general to be applicable to the study of soliton formation, all-optical switching and can be extended to processes involving other orders of the nonlinearity.\\
	
\begin{figure}
\centering
\includegraphics[height=1.5in]{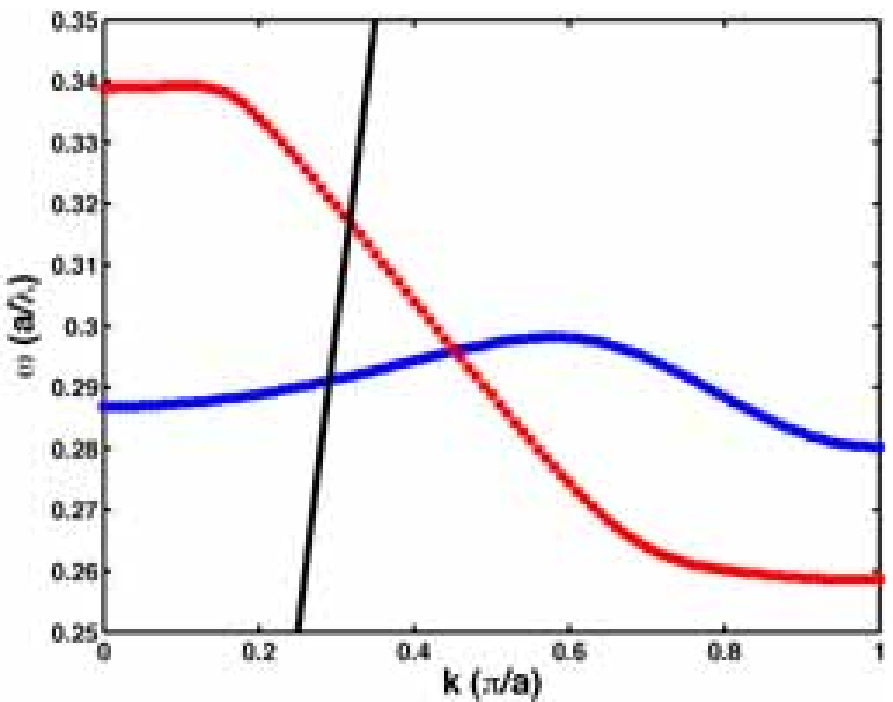}
\includegraphics[height=1.5in]{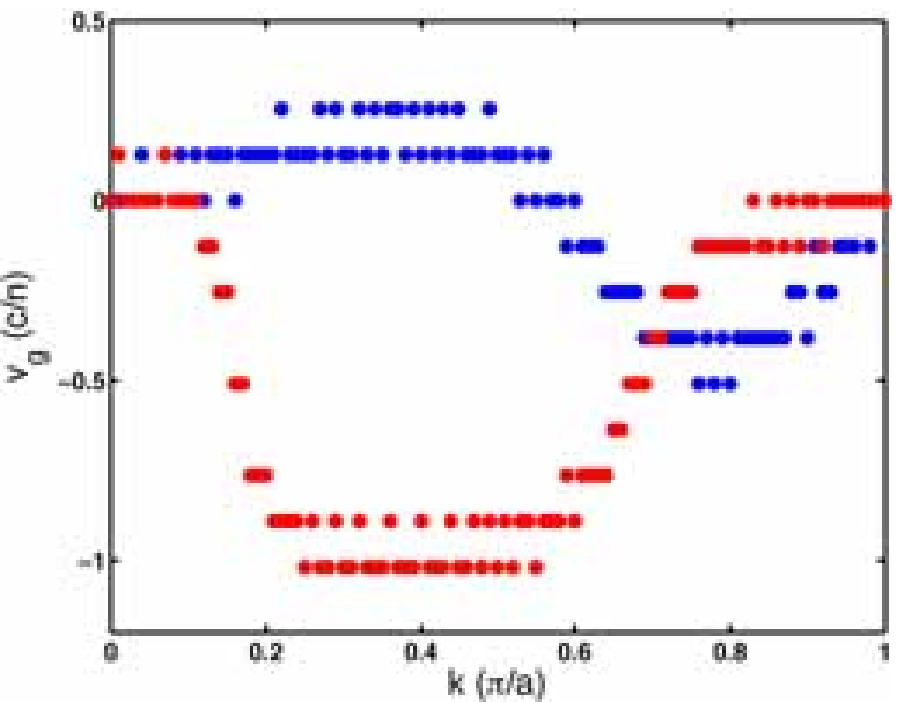} \\
\includegraphics[height=1.5in]{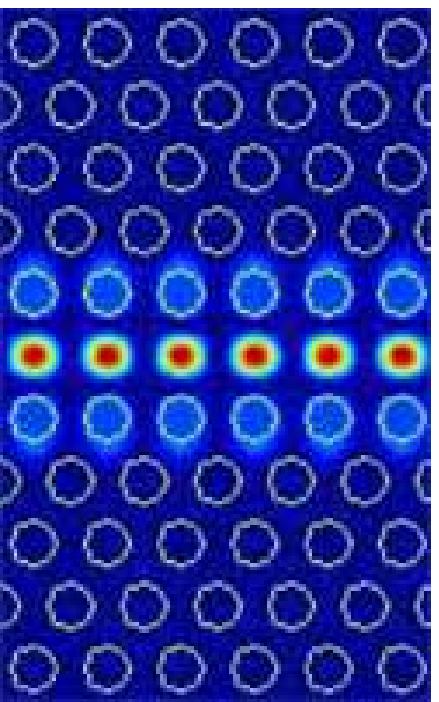}
\includegraphics[height=1.5in]{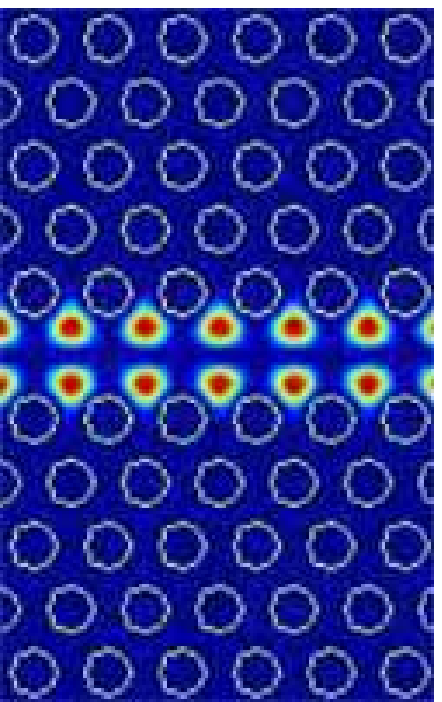}
\includegraphics[height=1.5in]{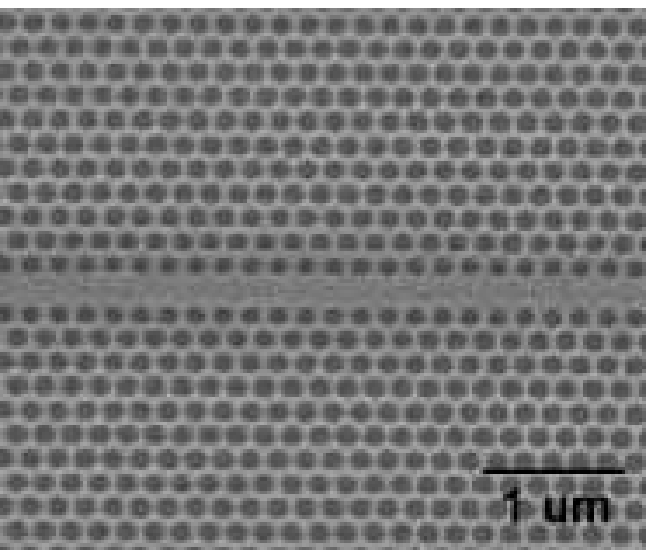}
	\caption{Top left: waveguide mode dispersions calculated by the 3D Finite Difference Time Domain (FDTD) method. The solid (black) line is the light line in the photonic crystal, above which modes are not confined by total internal reflection.Top right: group velocities of the two modes derived from the disperion curves by numerical differentiation $(v_g=\frac{d\omega}{dk})$. Bottom left: modes of the PhC lattice at the $k=\frac{\pi}{a}$ point. The mode with even vertical symmetry relative to the waveguide axis corresponds to the red dispersion curve and the odd symmetry mode is that of the blue dispersion curve. Bottom right: A scanning electron micrograph of a fabricated PCW in AlGaAs.}
	 \label{fig:wgmodes}
\end{figure}

	In the theoretical proposal, the probe is sent through two arms of a Michaelson interferometer and interacts with the signal in one arm. The phase shift on the probe is measured on the slope of the fringe via homodyne detection  \cite{ref:KerrY}. In our case, the interferometer would be made in a free standing membrane of AlGaAs that is patterned by a hexagonal lattice of air holes. The waveguides are made by removing rows of holes.  Fig. \ref{fig:wgmodes} shows a PCW in a hexagonal lattice and the PCW dispersion for two modes. The group velocity is significantly reduced at the band edge $(k_x=\frac{\pi}{a})$, which makes this an attractive operating point. Numerical precision allows us to estimate that $v_{g} < c\times 10^{-3}$. Since the Kerr effect depends on the intensity overlap, either two spectrally different points on the same PCW band, or on different bands can be chosen. In the first case, the intensity overlap is maximized, but there is a potential for a large group velocity mismatch. In the second case the mode overlap is sacrificed in favor of matching the group velocities.  In principle, $\frac{\Delta\omega}{\omega} \approx 10^{-6}$ for a $1 ns$ pulse, which means that ponts with very close $\frac{a}{\lambda}$ values can be chosen, and the proximity is limited by the ability to filter, or by the wavelength requirements for the pulse and probe.

First, we derive the equations of motion for the signal and probe pulse in the interferometer. The eigenstates of the PhC waveguide are solutions to:
\begin{equation}
\label{eq:Hw}
\nabla\times\nabla\times\vec{E}=-\frac{1}{c^2}\frac{\partial^2 (\epsilon (\vec{r})\vec{E})}{\partial{t}^2}
\end{equation}      
where $\epsilon (\vec{r})$ is the relative spatially varying waveguide dielectric constant. The solutions are Bloch modes $u^{m}_{k}(\vec{r})e^{i(kz-\omega(k)t)}$ where $u^{m}_{k}(\vec{r}+az)=u^{m}_{k}(\vec{r})$ for the lattice with periodicity a, z is the direction of propagation along the waveguide, and m is the index of the band of particular symmetry. These modes satisfy the wave equation:
\begin{equation}
\label{eq:BlochEq}
\nabla\times\nabla\times(u_{k}(\vec{r})e^{ikz})=\frac{\omega(k)^2}{c^2}\epsilon(\vec{r})u_{k}(\vec{r})e^{ikz}
\end{equation}      
	The Bloch state is re-normalized for convenience in the last part of the paper. The waveguide modes can be shown to obey the following orthogonality conditions  :

\begin{equation}
\label{eq:orthog1}
\int_{\Omega}{d^3r\epsilon(\vec{r}) u^{m}_{k}u^{n}_{k'}e^{i(k-k')z}}=\delta_{mn}\delta_{kk'} 
\end{equation}      
where the integral is taken over the whole space $\Omega$. In what follows, index m is dropped, unless it is necessary, and $\epsilon=\epsilon(r)$. The waveguide modes can be rewritten to solve a different Hermitian operator:
\begin{equation}
\label{eq:O}
\hat{O}=\frac{1}{\sqrt{\epsilon}}\nabla\times\nabla\times\frac{1}{\sqrt{\epsilon}}
\end{equation}      
The eigenstates of this operator are  $\langle{r}\ket{u,m,k} = \sqrt{\epsilon}u^{m}_{k}(\vec{r})e^{i(kz-\omega(k)t)}$ with eigenvalue $\frac{\omega(k)^2}{c^2}$, and $\braket{u,m,k}{u,n,l}=\delta_{m,n}\delta_{k,l}$ by Eq. \ref{eq:orthog1}. The inner product denotes integration over all physical space.

	A pulse propagating in the PhC waveguide in the presence of the weak nonlinearity may be written as $E=\frac{1}{\sqrt{\epsilon}}\int{dk A(k,t)\ket{u,m,k}}$ where $A(k,t)$ is a time dependent coefficient of each k component. The k-space range over which the integrand is appreciable depends on the frequency distribution of the pulse. For pulses with $\frac{\Delta\omega}{\omega} \approx 10^{-6}$, and group velocity of $v=\frac{c}{100}$, $\frac{\Delta k}{k} \approx 10^{-4}$. So the integrand in the expression for E is dominated by a particular k about which the pulse can be expanded:
\begin{equation}
\label{eq:envelope1}
E= \int{dk A(k,t) u_{k}(\vec{r})e^{i(kz-\omega(k)t)}} \approx u_{k_0}(\vec{r})e^{i(k_{0}z-\omega_{0}t)}\int{dk A(k,t)e^{i[(k-k_{0})z-(\omega(k)-\omega_{0})t]}} 
\end{equation}      
expanding $\omega(k)=\omega(k_0)+\frac{\partial\omega}{\partial{k}}|_{k=k_0}(k-k_{0})+\frac{1}{2}\frac{\partial^2\omega}{\partial{k}^2}|_{k=k_0}(k-k_{0})^2=\omega_{0}+v_{g}q+\frac{1}{2}\frac{\partial{v_{g}}}{\partial{}k}q^2$ with $q=k-k_{0}$ gives:

\begin{equation}
\label{eq:envelope2}
E \approx u_{k_0}(\vec{r})e^{i(k_{0}z-\omega_{0}t)}\int{ dq A(q+k_{0},t)e^{ i\left[q(z-v_{g}t)-\frac{1}{2}q^2\frac{\partial{v_{g}}}{\partial{}k}t\right]}}= u_{k_0}(\vec{r})e^{i(k_{0}z-\omega_{0}t)} \times F(z,t)=\frac{1}{\sqrt{\epsilon}}\ket{u,k_{0}} \times F(z,t)
\end{equation}     
	Here F is a slowly spatially and time varying envelope of the signal or probe, which extends over many periods of the waveguide. In order to determine the interaction of pulses propagating in the PhC waveguide, we need to know the evolution of such an envelope. In the Appendix, first order perturbation theory is applied to the operator $\hat{O}$ to determine the evolution of the Fourier components of the envelope. To first order in the nonlinear perturbation, and with negligible group velocity dispersion, the evolution of two pulses (S and P) in the same waveguide, but possibly coupled to different waveguide modes $(s,p)$ is given by (see Appendix):

\begin{align}
\label{eq:evS}
& \dot{S} = i \frac{1}{2} \kappa \omega_{s}(\gamma_{s,s}\nn{S}+2\gamma_{s,p}\nn{P})S-v_{s}S'+i\frac{1}{2}\frac{\partial{v_{s}}}{\partial{}k}S''\\
\label{eq:PevolApp}
& \dot{P} = i \frac{1}{2} \kappa \omega_{p}(\gamma_{p,p}\nn{P}+2\gamma_{p,s}\nn{S})P-v_{p}P'+i\frac{1}{2}\frac{\partial{v_{p}}}{\partial{}k}P''
\end{align}

with:
\begin{equation}
\label{eq:gamma}
\gamma_{s,p}= \frac{1}{a}\int_{\Lambda} d^3 r \tilde{\epsilon}(\vec{r})|u_{s}|^2 |u_{p}|^2
\end{equation}

	The dot in the above equation denotes differentiation in time, and the prime is a derivative in the direction of propagation (z). The overlap $\gamma_{s,p}$, has dimensions of $m^{-2}$, due to our re-normalization of the Bloch state to $\int_{\Lambda} d^3 r \nn{u}=a$. The function $\tilde{\epsilon}$ has a value of $n^{2}$ in the material and is zero in air. Here, p and s label the probe and signal modes at a particular k point, and $v_{s}$ and $v_{p}$ are the group velocities of the signal and probe pulses respectively; $\kappa$ is defined as $c \epsilon_0 n_2$ (see Appendix). The intensity of each pulse is enhanced by $\frac{c}{v_{p,s}}$, relative to bulk, as can be shown from Poynting's theorem \cite{ref:Jackson}. When there is no nonlinear coupling, each pulse propagates with group velocity $v_{s,(p)}$, and spreads according to $\frac{1}{2}\frac{\partial{v_{s,(p)}}}{\partial{k}}$. The above equations are derived in the Appendix and can be used to investigate self focusing, soliton formation, and other effects in PCW's. In the presense of the nonlinearity, the pulses experience self-phase modulation due to $\gamma_{s(p),s(p)}$ and cross-phase modulation due to $\gamma_{s,p}$ terms. The integral for $\gamma_{s,p}$, which gives the coupling strength, is taken over a unit cell of the waveguide, and is normalized by the length of the period a (see Appendix for further details). 

\begin{figure}
\centering
\includegraphics[height=1.5in]{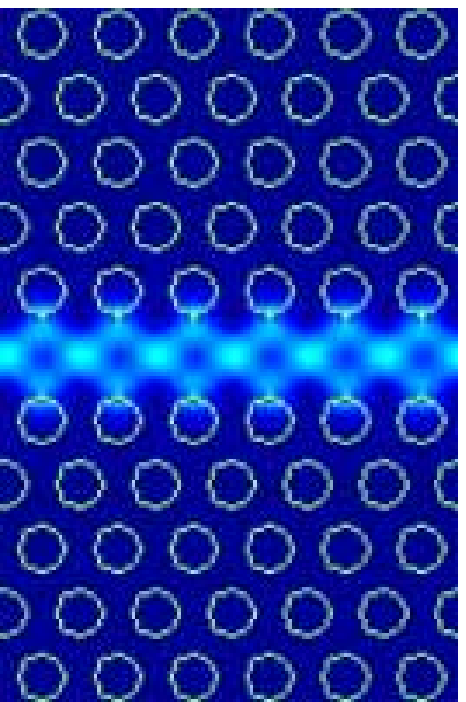}
\includegraphics[height=1.5in]{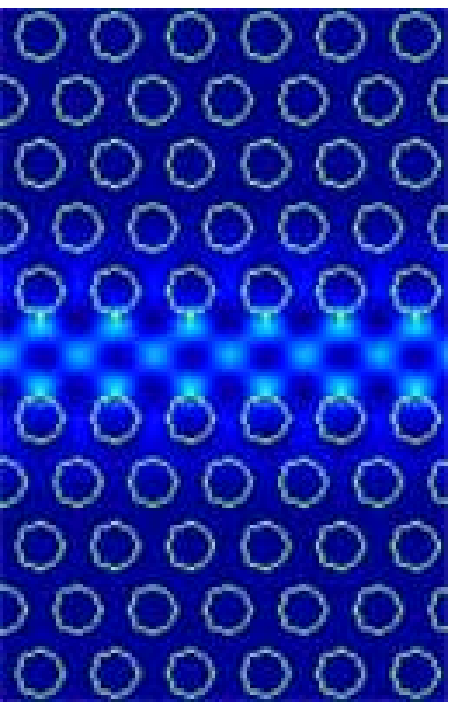}
\includegraphics[height=1.5in]{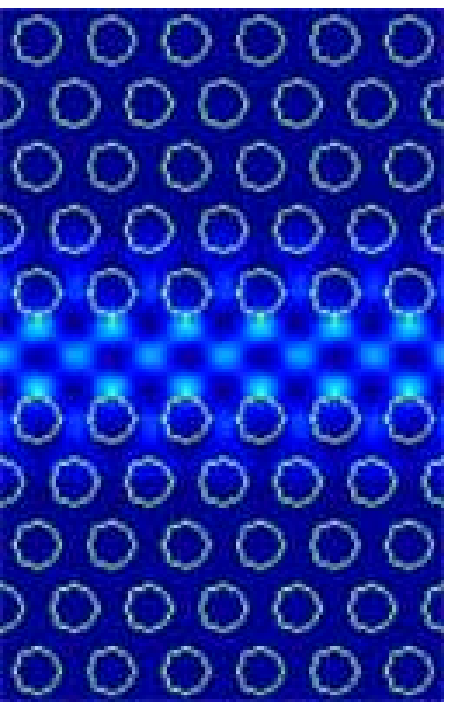}
\caption{Amplitude of E field for k = $ \frac{2}{3} \frac{\pi}{a}, \frac{5}{6} \frac{\pi}{a}$ and $\frac{\pi}{a}$}
\label{fig:wg_k}
\end{figure}

	The shape of $u_{s}$ and $u_{p}$, and hence the values of the $\gamma$ terms, is not strongly k dependent within numerical error for a wide range of wavevectors, as determined by 3D Finite Difference Time Domain (FDTD) simulations (Figure \ref{fig:wg_k}), and $\partial_{k} u_{k} \approx 0$. Thus, the coupling strength $\gamma_{s,p}$ only depends on the waveguide branch for the modes and not the particular k point. The total effective interaction strength is k dependent, since the group velocity determines the propagation time. The coupling strengths in units of $a^{-2}$, and mode volumes of the unit cell of the waveguide in units of $a^{-3}$, are shown in the Table \ref{table:Overlaps}. The mode volume for each waveguide mode is defined as $V_{1,1 (2,2)}=(\epsilon \nn{u_{1(2)}})^{-1}_{max}\int_{\Lambda} d^3 r \epsilon \nn{u_{1(2)}}$, and the mode volume for the overlap is $V_{1,2}=(\epsilon  |u_{1}||u_{1}|)^{-1}_{max}\int_{\Lambda} d^3 r \epsilon |u_{1}||u_{1}|$.

\begin{table}
	\caption{values for coupling $\gamma$ for different modes of the waveguide in units of $\frac{1}{a^2}$, and mode volumes for each unit cell of the waveguide (in units of $\frac{1}{a^3}$. $1$ and $2$ refer to the first and second modes of the waveguide.}
	 \label{table:Overlaps}
	 \begin{center}
          \begin{tabular}{|c|c|c|c|}
             \hline
               $u_{i,j}$ & $1,1$ & $2,2$ & $1,2$ \\
	 \hline
 	$\gamma_{i,j}$ & $ 6.4\times 10^{-2} $ & $7.9 \times 10^{-2} $ & $1.4\times 10^{-2} $\\
	\hline
 	$V_{mode}$ & $3.9 \times 10^{-1}$ &$ 2.8 \times 10^{-1}$ & $2.5\times 10^{-1} $\\
          \hline
          \end{tabular}
	\end{center}
\end{table}

	Each equation can be transformed into a coordinate frame moving with the probe and signal respectively via $x=z-v_{s}t$ and $x=z-v_{p}t$. The dispersion terms in \ref{eq:evS} complicate the solution. We will assume that the length of the waveguide is small enough so that the measurement of the induced phase and the measurement of the phase on the probe is unaffected by the dispersion throughout the propagation. With $\frac{\partial{v_{s}}}{\partial{}k}$ and $\frac{\partial{v_{p}}}{\partial{}k}$ neglected, the solution and upper bounds on the phases on the probe after time $t=\frac{L}{v_{p}}$ are:
\begin{align}
\label{eq:Formal_P_Solution}
P(z') &= Exp[-i \kappa \omega_{p} \int^{t}_{0}(\gamma_{p,p}\nn{P(z')}+2\gamma_{s,p}\nn{S(z'+\Delta{v}t')})dt']P(0)=P(0)e^{i(\phi_{P}+\phi_{S})} \\
\phi_{P} &=\frac{1}{2}\kappa \omega_{p} \int^{t}_{0}\gamma_{p,p}\nn{P(z')}dt' \approx \frac{1}{2} \kappa \omega_{p}\gamma_{p,p} \frac{L}{v} \nn{P(z')} \\
\phi_{S} &=\kappa \omega_{p} \int^{t}_{0}\gamma_{s,p}\nn{S(z'+\Delta{v}t')}dt'  \approx  \kappa \omega_{p}\gamma_{s,p} \frac{L}{v} \nn{S(z')} \\
\end{align}

	The phase $\phi_{S}$ is the phase on the probe due to the nonlinear interaction with the signal, and gives the signal photon number.  In the appendix we derive that the ideal case of a negligible group velocity mismatch and a narrow probe, gives the phase shift per signal photon of:
\begin{equation}
\label{eq:phase_ideal}
\phi_{s,ideal} = c n_{2} \gamma_{s,p} \hbar \omega_{s} \omega_{p} \frac{L}{v_{p}}\frac{1}{v_{s} \tau_{s}}
\end{equation}
L is the length of the PhC and $\tau_{s}$ is the temporal width of the signal wavepacket. \\

	Both the signal and probe wave undergo material absorption and scattering due to waveguide losses. PCW losses are already below $1 dB/\mu m$ and will improve with time. The material absorption consists of the linear absorption coefficient $\alpha_{1}$ and the nonlinear coefficients, of which we will only consider the two photon absorption coefficient $\alpha_{2}$. In the case of AlGaAs at the half bandgap, the values of $\alpha_{1}$ and $\alpha_{2}$ were found to be $\approx .1 cm^{-1}$ and $\approx .2 \frac{cm}{GW}$ respectively \cite{ref:Aitchison}. Thus, $\alpha_{1}$ limits us to $L\frac{c}{v_{p}} \approx 10 cm$. For $\mu$J and sub-$\mu$J pulses, $\alpha_{2}$ results in a smaller attenuation length on the order of $50 \mu m$ at best.  Thus, experiments with the signal and probe at the half-gap are not feasible. In order to circumvent pump depletion due to two photon absorption, a pump at even longer wavelengths above 1550 nm should be used \cite{ref:AlGaAs_Villeneuve}. In that case, $\alpha_{2}$ is close to zero, and we will assume that the 100 $\mu m$  PCW length is the limit. In this case, the pump will propagate in the lower  branch of the waveguide, while the signal should couple to the upper branch. For example for a signal at 1550 nm in the upper waveguide branch, a pump at 1620 nm should be used in order to have both beams velocity matched at the $\pi/a$ point. We briefly mention that the GaInAsP material system has $\alpha_{1}\approx 1 cm^{-1}$, and $n_{2} \approx 5 \times 10^{-12} \frac{cm^2}{W}$ and most likely similar two-photon absorption, which means that both materials are suitable candidates for an experiment. While the nonlinearity is enhanced closer to the band-edge of the semiconductor bandgap, the absorption increases accordingly and reduced the interaction length. \\

\begin{figure}
\centering
\includegraphics[height=1 in] {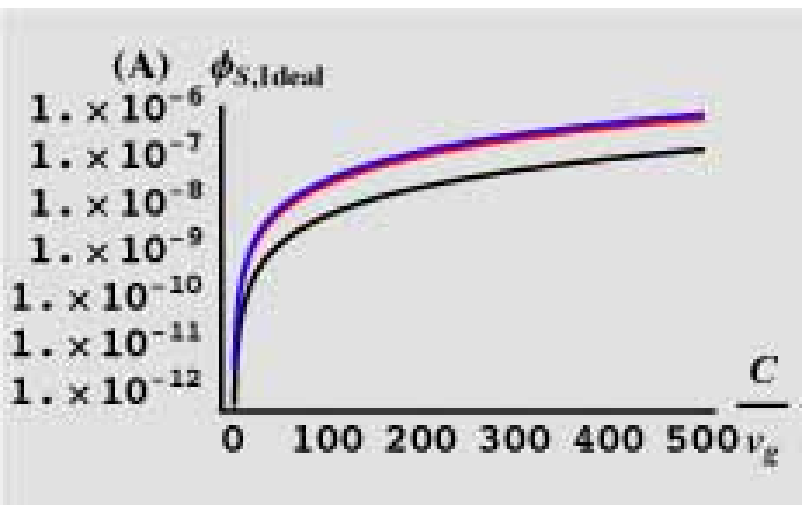}
\includegraphics[height=1 in] {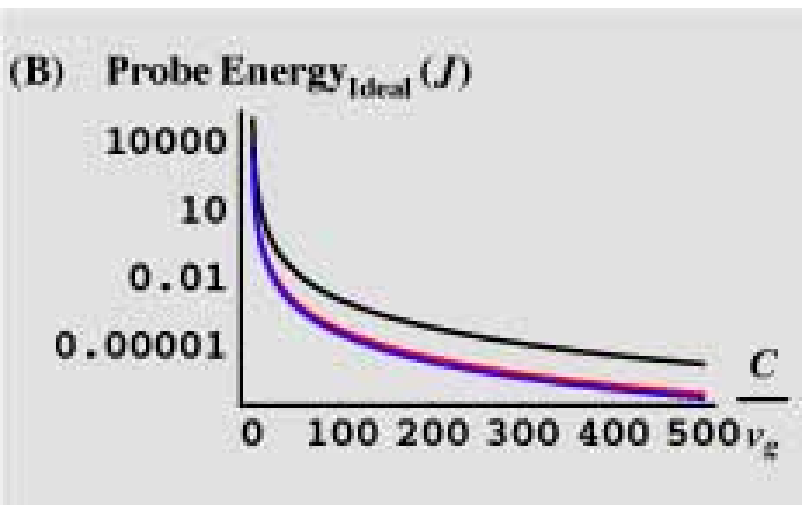} \\
\includegraphics[height=1 in] {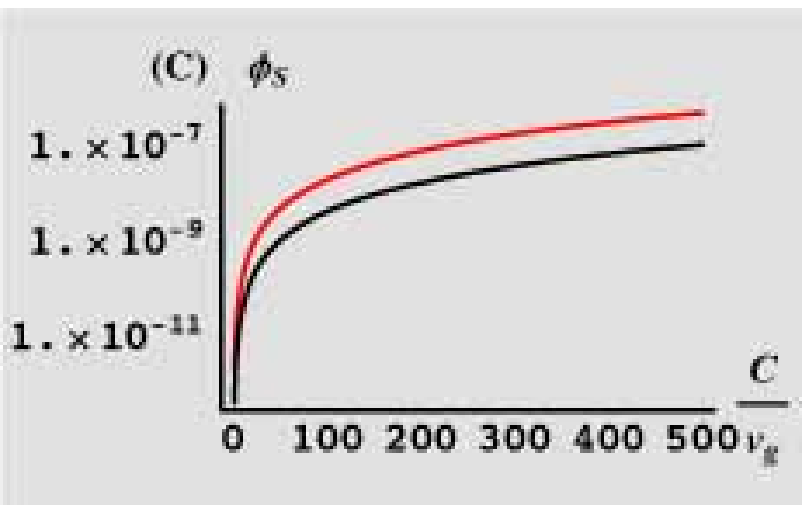}
\includegraphics[height=1 in] {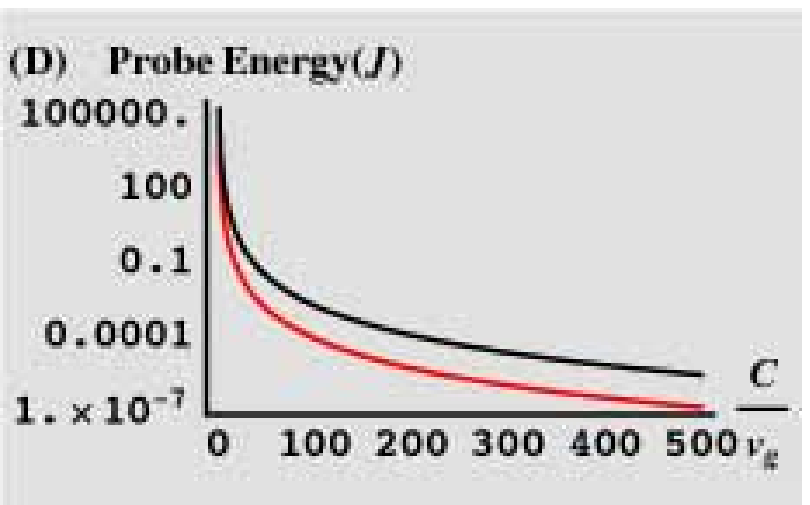}
\caption{Phase shift due to a single signal photon with a lifetime of 200 ps, after propagation through a 100 $\mu m$ AlGaAs waveguide with a narrow probe and no group velocity mismatch (A), and the energy required for an external pulse to obtain a SNR of 1 (B). In (A) and (B) it is assumed that the signal and probe are at 1500 nm and two-photon absorption is not present. In (C) we plot the phase for the case of the signal photon in waveguide 1 at 1550 nm and probe at 1620 nm in waveguide 0. The required probe energy for this scheme is shown in (D). In all plots, the blue and red curves correspond to both the signal and the probe in waveguide modes 0 or 1. The black curve corresponds to the probe and signal in different waveguide modes} 
\label{fig:phase_vg}
\end{figure}

	The phase due to a single photon in signal S and the energy required for an SNR of 1 for number state detection in AlGaAs, are plotted in Fig. \ref{fig:phase_vg}. We plot both the ideal case, in which two-photon absorption is negligible, and the reality in which the pump is at $1620 nm$. 

	There are two sources of noise in this experiment in the case of an ideal detector. One is the phase noise due to intrinsic noise of the signal beam, and the other is the interferometer  noise due to the uncertainty of the probe photon number. Following \cite{ref:MQO}, it can be shown that in the case of a  coherent signal state with mean photon number $\langle \hat{n}_{s}\rangle=N_{s}$ and coherent probe with mean photon number $\langle \hat{n}_{p}\rangle=N_{s}$ the uncertainty in the detected signal is $\langle \Delta n^{2}_{s,observed} \rangle =\langle \Delta n^{2}_{s,intrinsic} \rangle+\frac{\langle \Delta \hat{n}^{2}_{p} \rangle}{\phi^{2}_{s} N^{2}_{p}}$. There are two cases of interest: the signal in coherent and number states. For the coherent state,
\begin{equation}
\label{eq:coherent_noise}
\frac{1}{\langle \Delta \hat{n}^{2}_{s,observed} \rangle}= \frac{1}{N_{s}}+ 4 \phi^{2}_{s} N_{p}
\end{equation}
In the case of the signal in the number state, the intrinsic noise of the signal disappears,
\begin{equation}
\label{eq:number_noise}
\langle \Delta \hat{n}^{2}_{s,observed} \rangle =\frac{1}{4 \phi^{2}_{s} N_{p}}
\end{equation}
When the probe photon number is reasonably large, we can relax the requirement on $N_p$. If the  tolerated error for coherent state detection is $E=\beta N_s$, then the condition is $N_p = (4 \phi^{2}_{s} \beta N_{s})^{-1}$, and $\beta < 1$. Thus, detection of $1000$ signal photons with an error of $100$ ($\beta=0.1$), would require $50-100 nJ$. For smaller signal photon numbers, the level of tolerated error decreases, and the requirement is more stringent than number state detection, since $\frac{\beta}{N_s} > 1$. \\

	In conclusion, we have derived the equations of motion for a probe and signal wave interacting via the third order nonlinearity in a photonic crystal waveguide. Within the slowly varying envelope approximation, the equations yield intuitive results, and are essentially identical to the equations of propagation for pulses in nonlinear fibers and materials, if the plane waves used in the mode expansion of the electromagnetic fields are replaced by Bloch waves. However, the user of PCW leads to the necessary enhancement of the pulse intensities due to the small mode volume and reduced group velocity of the pulses.  We have shown that for the case of a very long wavelength probe pulse, that does not suffer from two-photon absorption in the AlGaAs material system, the energy requirement on the probe wave is within attainable values ($\approx \mu J$ in sub $ns$ pulses). Since the sources of such pulses are external to the PCW, the generated probe pulse can be broader than the pulse desired in the PCW, due to contraction by the group velocity. Our derivation has assumed that coupling into the waveguides and the beamsplitter implementation in a PCW\ are perfect, and the scattering loss of the waveguide can be neglected. This is of course a gross generalization. High coupling efficiencies and low loss propagation over 10's of microns have been shown, and will only improve in time. We are in general most strongly limited by two photon absorption in this proposal. Other material systems such as GaInAsP \cite{ref:all_optical_GaInAsP}, which also exhibit higher $n_{2}$ values that AlGaAs, could also be considered for this implementation, but ultimately atomic resonance systems and systems with a high phase shift and low loss are necessary \cite{ref:Edo_Kerr,ref:HP_Kerr}. In principle, an on-chip QND photon number detector could be a component of a photonic crystal based quantum circuit, or can serve as a sensitive intensity detector and switch. The derivation presented here, can be easily extended to other types of intensity and field dependent nonlinearities, and can be used to analyze other nonlinear optical effects in PCW's, as well as soliton formation and propagation.

\acknowledgements{
We would like to thank Dr. Edo Waks for his insight and help with the preparation of this manuscript. Financial support was provided by the MURI Center for photonic quantum information systems (ARO/ARDA
Program DAAD19-03-1-0199) and the DARPA Nanophotonics seed grant. Ilya Fushman is supported by the NDSEG fellowship}

\appendix
\section{Derivation of the propagation equation}

The probe envelope evolves according to (with $q=k-k_0$):
\begin{align}
\frac{\partial{P}}{\partial{t}} &=\frac{\partial}{\partial{t}}\int{ dk A(k,t)e^{ i\left[(k-k_0)z-(\omega(k)-\omega_0)t\right]}} \\
 & \approx \frac{\partial}{\partial{t}}\int{ dq A(q+k_{0},t)e^{ i\left[q(z-v_{g}t)-\frac{1}{2}q^2\frac{\partial{v_{g}}}{\partial{}k}t\right]}} \\
&= \int{ dq (\frac{\partial{A(q+k_{0},t)}}{\partial{t}}-i(v_{g}q+\frac{1}{2}\frac{\partial{v_{g}}}{\partial{}k}q^2) A(q+k_{0},t))e^{ i\left[q(z-v_{g}t)-\frac{1}{2}q^2\frac{\partial{v_{g}}}{\partial{}k}t\right]}} \\
&= \int{ dq \frac{\partial{(A(q+k_{0},t)}}{\partial{t}}e^{i\left[q(z-v_{g}t)-\frac{1}{2}q^2\frac{\partial{v_{g}}}{\partial{}k}t\right]}}-(v_{g}\frac{\partial}{\partial{z}}-i\frac{1}{2}\frac{\partial{v_{g}}}{\partial{}k}\frac{\partial^2}{\partial{z}^2})\int{dq A(q+k_{0},t))e^{i\left[q(z-v_{g}t)-\frac{1}{2}q^2\frac{\partial{v_{g}}}{\partial{}k}t\right]}} \\
\label{eq:dS}
&=\int{ dq \frac{\partial{A(q+k_{0},t)}}{\partial{t}} e^{i\left[q(z-v_{g}t)-\frac{1}{2}q^2\frac{\partial{v_{g}}}{\partial{}k}t\right]}}-v_{g}\frac{\partial}{\partial{z}}P+i\frac{1}{2}\frac{\partial{v_{g}}}{\partial{}k}\frac{\partial^2}{\partial{z}^2}P
\end{align}
 in the case of nonlinearity, the wave equation with $c^{-2}=\mu_{0}\epsilon_{0}$ is:
\begin{equation}
\label{eq:MaxNonlinear}
-\nabla\times\nabla\times\vec{E}=\frac{1}{c^2}\frac{\partial^2 (\epsilon(\vec{r})\vec{E})}{\partial{t}^2}+\mu_{0}\frac{\partial^2\vec{P}}{\partial{t}^2}
\end{equation}  
And we can rewrite this equation in terms of the previously introduced Hermitian field operator as:
\begin{equation}
\label{eq:MaxNonlinear}
-\frac{1}{\sqrt{\epsilon}}\nabla\times\nabla\times\frac{1}{\sqrt{\epsilon}}\sqrt{\epsilon} \vec{E}=-\hat{O}\sqrt{\epsilon}\vec{E}=\frac{1}{c^2}\frac{\partial^2 (\sqrt{\epsilon}\vec{E})}{\partial{t}^2}+\mu_{0}\frac{\partial^2}{\partial{t}^2}\frac{1}{\sqrt{\epsilon}}\vec{P}
\end{equation}

	Here, in general, the polarizability P is given by $P_{i}=\epsilon_{0} \sum_{j,k,l}\sum_{m,n,p}\chi^{(3)}_{i,j,k,l}E(\omega_{p})_{i}E(\omega_{s})_{j}E(\omega_{p})_{k}$, where $\chi^{(3)}_{i,j,k,l}$ are the components of the third order nonlinearity tensor. In our case, we consider the system to be isotropic, the response instantaneous, and only two frequencies ($\omega_s,\omega_p$) to be present.
To find the time evolution of the coefficients, we use the wave equation. In the presence of the nonlinearity we expand $\epsilon \approx \epsilon+\delta$ and $\hat{O}$ as:
\begin{align}
\label{eq:Oexpand}
\hat{O}+\Delta\hat{{O}} &=\frac{1}{\sqrt{\epsilon}}\nabla\times\nabla\times\frac{1}{\sqrt{\epsilon}}-\frac{1}{2}\left[
\frac{\delta}{\epsilon}\frac{1}{\sqrt{\epsilon}}\nabla\times\nabla\times\frac{1}{\sqrt{\epsilon}}+
\frac{1}{\sqrt{\epsilon}}\nabla\times\nabla\times\frac{1}{\sqrt{\epsilon}}\frac{\delta}{\epsilon}
\right]+o\left[\left(\frac{\delta}{\epsilon}\right)^{2}\right]\\
\hat{O}+\Delta\hat{{O}} &\approx \hat{O}-\frac{1}{2}[
\frac{\delta}{\epsilon}\hat{O}+
\hat{O}\frac{\delta}{\epsilon}]= \hat{O}-\frac{1}{2}\{
\frac{\delta}{\epsilon},\hat{O}\}
\end{align}
Let $\bra{u'}=\bra{u,k',n'}$, $\ket{u}=\ket{u,k,n}$, $A=A(k,t)$. Then we have:
\begin{align}
\bra{u'}(\hat{O}+\Delta\hat{{O}})\int{dk A\ket{u}} &= -\bra{u'}\frac{1}{c^2}\frac{\partial{}^2}{\partial{t}^2}\int{dk A\ket{u}} \\
\int{dk A (\bra{u'}\hat{O}\ket{u}+\bra{u'}\Delta\hat{{O}}) \ket{u}} &=- \bra{u'}\frac{1}{c^2}\int{dk (\ddot{A}-\omega^{2}A-2i\omega{\dot{A}})\ket{u}} \\
\int{dk A \bra{u'}\Delta\hat{{O}} \ket{u}} &= \frac{2i\omega}{c^2}\int{dk{\dot{A}} \braket{u'}{u}} \\
-\frac{1}{2}\int{dk A \bra{u'}
\frac{\delta}{\epsilon}\hat{O}+
\hat{O}\frac{\delta}{\epsilon}\ket{u}} &= \frac{2i\omega}{c^2}\dot{A}(k')\\
-\frac{1}{2}\int{dk} A \bra{u'}\frac{\delta}{\epsilon}\ket{u}(\frac{\omega^{'2}}{c^2}+\frac{\omega^{2}}{c^2}) &= \frac{2i\omega}{c^2}\dot{A}(k')\\
\label{eq:Ak_time_evol}
-\frac{\omega^{2}}{c^2}\int{dk A(k) \bra{u'}\frac{\delta}{\epsilon}\ket{u}} &= \frac{2i\omega}{c^2}\dot{A}(k')
\end{align}
 	Above, we neglect second derivatives of the envelope and combine the two frequency terms as $\omega^{'2}+\omega^{2} \approx 2 \omega^{2}+v_{s}(k-k') \approx 2\omega^{2}$, because the $(k-k')$ term will lead to the derivative of the slowly varying envelope multiplied by the nonlinearity and is very small. Since the frequency bandwidth of the envelope is small, the envelope is slowly varying in time, and the second order time variation in the coefficients $A(k)$ is neglected. Furthermore, we have also assumed that the inner product $\braket{u'}{u}$ is roughly unchanged by the nonlinearity -- it remains a delta function. The perturbation $\delta_{s,p}$ contains both the real and imaginary parts of the third order susceptibility. The real part is responsible for the cross phase modulation, while the nonlinear term gives the two photon absorption of the signal and probe. The figure of merit for the feasibility of the experiments is the phase-shift gained per loss length ($\frac{1}{e}$ point). The full perturbation can be written as:
\begin{align}
\label{eq:perturbation_full}
  \delta_{s,p}=i\alpha_{1}+ 3\epsilon_{0}(\chi^{(3)}_{r}+i\chi^{(3)}_{i})(|E(\omega_{s,p})|^{2}+2|E(\omega_{p,s})|^{2}) 
\end{align}
 	Where $\chi^{(3)}$ is the third order polarizability (which is assumed to have only one value and to be infinitely fast), and $E(\omega_{s,p})=\{S,P\}u_{k_{s,p}}e^{i (k_{s,p} z- \omega_{s,p} t)}$ is the electric field of the two modes. The linear loss $\alpha_{1}$ is the imaginary part of the dielectric constant.
\\
	The value of $\chi^{(3)}_{r}$ can be determined from the experimentally observed bulk intensity dependent refractive index defined via $\tilde{n}=n + n_{2} I$, where I is the average field intensity. For a bulk material I is given by $I=\frac{1}{2} \sqrt{\frac{\epsilon_0 \epsilon}{\mu_0}}\nn{E}=\frac{1}{2} c\epsilon_{0} n \nn{E}$. Thus,  $3 \chi^{(3)}= c n^2 n_{2}$. For AlGaAs at the wavelength of 1.5 $\mu m$, $n_{2} \approx 1.5 \times 10^{-13} \frac{cm^2}{W}$; furthermore, the index is very similar for TE and TM polarization in AlGaAs slab waveguides \cite{ref:Aitchison}. Since $\chi^{(3)}$  measures the response of the local charge distribution to the local electric field, the coefficient itself is a material property and is not modified in the PCW, except possibly due to surface effects (e.g. reduced response or artificially added birefringence). Thus, we can derive the value of the coefficient from bulk experiments and combine it with the modified electromagnetic fields to get the resulting effect in the PCW. The $\chi^{(3)}$ coupling term only exists in the material, and we can replace the $n^{2}$ term with a dielectric which is equal to the spatially patterned index of AlGaAs in the PCW and is zero in the air. We set $3 \chi^{(3)} = c n_{2}\tilde{ \epsilon}(\vec{r})$. And we define $\kappa= c \epsilon_{0} n_2$, so that the perturbation due to real part of the nonlinearity becomes $\delta_{s,p}=\kappa \tilde{\epsilon}(\vec{r}) (|E(\omega_{s,p})|^{2}+2|E(\omega_{p,s})|^{2})$ . The loss terms are similarly determined from a fit to $\alpha_{total}=\alpha_{1}+\alpha_{2} I$. The linear loss $\alpha_{1}$ results in an exponential decay of the signal with a characteristic length $(\alpha_{1})^{-1}$. The nonlinear loss gives a characteristic length of $(\alpha_{2} I)^{-1}$. We will drop the losses for now, in order to derive the equation of motion for the pulses, and will assume that the Bloch components of the eigenstate $\ket{u}$ and $\ket{u'}$ belong to the same waveguide branch $n=n'$. Each of $u_{n,k}$, $u_{n,k'}$ is then roughly given by some central k component that is modulated by an envelope $u_{n,k}e^{ikz}\approx u_{n,k_{0}}e^{i(k_{0}z-\omega_{0}t)}e^{i[(k-k_{0})z-(\omega(k)-\omega_{0})t]}$. We now insert the exact form for $\bra{u'}$ and $\ket{u}$ into the above equation, and only look at the cross phase modulation component on the probe due to the signal. We take n to be branch of the pump mode p, and m to be that of the signal s, and only treat the perturbation due to the signal explicitly. Also, we will assume for simplicity that the signal group velocity $v_{s}$ at k and k', as well as the dispersion $\frac{\partial{v_{s}}}{\partial{}k}$, so that we can combine them in the expansion of the Bloch state. Eq. \ref{eq:Ak_time_evol} then implies:
\begin{align}
\dot{A}(k') &= i \omega_{p} \int{dk''} \int{d^3 r'} A(k'') \kappa \tilde{\epsilon} \nn{u_{p}} \nn{u_{s}} \nn{S(z')} e^{i[(k''-k')z'-(\omega(k'')-\omega(k'))t]} \\
 & \approx   i \omega_{p} \kappa  \int{dk''}A(k'') \int{dz'} \nn{S(z')}e^{i[(k''-k')z'-(\omega(k'')-\omega(k'))t]} \frac{1}{a} \int_{\Lambda}{d
 ^3 r}\tilde{\epsilon} \nn{u_{s}}\nn{u_{p}} \\	
  &\approx   i \omega_{p} \kappa  \gamma_{s,p} \int{dk''}A(k'') \int{dz'} \nn{S(z')}e^{i[(k''-k')z'-(\omega(k'')-\omega(k'))t]} 
\end{align}

	This is now inserted back into the evolution equation for the envelope \ref{eq:dS}, and we re-substitute $k-k_{0}=q$ and keep the frequency term in the form $\omega(k)-\omega(k_0)$ for convenience.
\begin{align}
& \int{dk}\dot{A}(k,t)e^{i\left(k-k_0)z-(\omega(k)-\omega_0)t\right)} = \\
&=  i \omega_{p} \kappa \gamma_{s,p}  \int dk \int dk'' A(k'') \int dz' \nn{S(z')} e^{i[(k''-k)z' - (\omega(k'')-\omega(k))t]} e^{i[(k-k_0)z-(\omega(k)-\omega_{0})t]}\\
&=  i \omega_{p} \kappa  \gamma_{s,p}  \int dz' \int dk'' A(k'') \nn{S(z')} e^{i[(k'' z'-k_0 z)-(\omega(k'')-\omega_{0})t])} \int dk e^{i k (z-z')}\\
&=  i \omega_{p} \kappa  \gamma_{s,p}  \int dz' \nn{S(z')} e^{i k_0(z'- z) } \delta(z-z') \int dk'' A(k'') e^{i[(k''-k_0) z'-(\omega(k'')-\omega_{0})t])}\\
&=  i \omega_{p} \kappa  \gamma_{s,p}  \int dz' \nn{S(z')} e^{i k_0(z'- z) } \delta(z-z') P(z')\\
&=  i \omega_{p} \kappa  \gamma_{s,p}  \nn{S(z)}P(z)
\end{align}

	The $\sqrt{\epsilon}$ terms cancel the denominator of the perturbation. The term $\gamma_{s,p}$ contains an effective area integral $\gamma_{s,p}= \frac{1}{a} \int_{\Lambda}{d
 ^3 r}\tilde{\epsilon} \nn{u_{s}}\nn{u_{p}} \approx \int{dx dy} \tilde{\epsilon} \nn{u_{s}}\nn{u_{p}}$. Since the Bloch states are periodic, their integral in each unit cell ($\gamma_{s,p}$) is the same, and we simply weigh it by the average value of the slowly varying envelope in that cell to find the integral over the whole volume. $\Lambda$ is the unit cell volume. S and P are only functions of time and the propagation coordinate (z here), and are uniform in the transverse (x,y) plane. The term $ \kappa $ contains the strength of the nonlinearity. We now re-normalize the Bloch state and the field:

\begin{align}
N_{s} \hbar \omega_{s} &=\int\int\int d^3\vec{r} \epsilon_{0} \epsilon(\vec{r}) \nn{S}\nn{u_s}\\
& \approx \int dz \epsilon_{0} \nn{S} \frac{1}{a} \int_{\Lambda} d^3 \vec{r} \epsilon(\vec{r}) \nn{u_{s}} \\
N_{s} \hbar \omega_{s} & = \int dz \epsilon_{0} \nn{S} \\
1 &=  \frac{1}{a} \int_{\Lambda} d^3 \vec{r} \epsilon(\vec{r}) \nn{u_{s}} 
\end{align}

	The above normalization means that $[\gamma]=m^{-2}$ and $[\nn{S}]=Volt^2$. Thus the term $\kappa \gamma \omega \nn{S}$ has units of $s^{-1}$, as desired. We now insert the form for $\dot{A}$ into \ref{eq:dS}: 

	The refractive index due to cross-phase modulation is twice that of self phase modulation \cite{ref:Boyd}. The evolution of the slowly varying envelope is given by:
\begin{align}
\label{eq:SevolApp}
& \dot{S} = i \frac{1}{2} \kappa \omega_{s}(\gamma_{s,s}\nn{S}+2\gamma_{s,p}\nn{P})S-v_{s}S'+i\frac{1}{2}\frac{\partial{v_{s}}}{\partial{}k}S''\\
\label{eq:PevolApp}
& \dot{P} = i \frac{1}{2} \kappa \omega_{p}(\gamma_{p,p}\nn{P}+2\gamma_{p,s}\nn{S})P-v_{p}P'+i\frac{1}{2}\frac{\partial{v_{p}}}{\partial{}k}P''
\end{align}

	We will now show the qualitative behavior of the two pulses, assuming a weak interaction. Take $P = \rho(z,t)e^{i\phi(z,t)}$ and $S = \eta(z,t)e^{i\psi(z,t)}$. Inserting P into \ref{eq:PevolApp}, the real and imaginary parts satisfy: \\
Real:
\begin{align}
\label{eq:rho_real}
& \dot{\rho}+(v_{p}+\frac{\partial{v_{p}}}{\partial{}k} \phi^{'})\rho^{'}=-\frac{\frac{\partial{v_{p}}}{\partial{}k}}{2}\phi^{''}\rho
\end{align}
Imaginary:
\begin{align}
\label{eq:rho_imag}
\dot{\phi}\rho+v_{p}\phi^{'}\rho= \frac{1}{2} \kappa \omega_{p}(\gamma_{p,p}\rho^{2}+2\gamma_{s,p}\eta^{2})\rho+\frac{\frac{\partial{v_{p}}}{\partial{}k}}{2}(\rho^{''}-(\phi^{'})^{2}\rho)
\end{align}

	If we assume that in \ref{eq:rho_real} the second derivative term vanishes, then we can see that the envelope moves along a characteristic given by $v_{p} + \frac{\partial{v_{p}}}{\partial{}k} \phi^{'} \approx v_{p}(k_{0}+ \phi^{'})$, which, in the flatter regions of the dispersion curve is very close to $v_{p}(k_0)$. If $\frac{\partial{v_{p}}}{\partial{}k} (\frac{\rho^{''}}{\rho}  )<< 1$, then \ref{eq:rho_imag} simplifies to give an equation for the phase along a characteristic given by $v_{p}(k_0+\frac{\phi^{'}}{2})$. 
\begin{align}
\dot{\phi}+v_{p}\phi^{'}+\frac{\frac{\partial{v_{p}}}{\partial{}k}}{2}(\phi^{'})^{2} = \frac{1}{2} \kappa \omega_{p}(\gamma_{p,p}\rho^{2}+2\gamma_{s,p}\eta^{2})
\end{align}

	 Since we will not generally be able to generate a pulse that is a solution to the nonlinear system, we will assume that the pulses are Gaussian and drop the dispersion terms for simplicity of the analysis. We are essentially assuming that the envelopes are very slowly varying and that the interaction time and dispersion are not strong enough to affect the phase measurement, which is dominated by the group velocity term rather than the group velocity dispersion term. Thus, we will rewrite Eq. \ref{eq:PevolApp} in a frame moving with the group velocity in terms of $z'=z-v_{p}t$ and $t'=t$, and neglect the terms with $\frac{\partial{v_{p}}}{\partial{}k}$ and $\frac{\partial{v_{s}}}{\partial{}k}$. Since the two pulses may have different group velocities, we have $\Delta{v}=(v_{p}-v_{s})$. If the group velocity dispersion terms are neglected, each envelope is only a function of $z'$. 
\begin{align}
\label{eq:Pe_Final_Equation}
& \frac{d{P(z')}}{dt'} \approx \frac{i}{2} \kappa \omega_{p}(\gamma_{p,p}\nn{P(z')}+2\gamma_{s,p}\nn{S(z'+\Delta{v}t)})P(z')
\end{align}
The solution, and upper bounds on the phases after time $t=\frac{L}{v_p}$ are:
\begin{align}
\label{eq:Formal_P_Solution}
P(z') &=P(0) Exp[\frac{i}{2}  \kappa \omega_{p} \int^{t}_{0}(\gamma_{p,p}\nn{P(z')}+2\gamma_{s,p}\nn{S(z'+\Delta{v}t')})dt']=P(0)e^{i(\phi_{P}+\phi_{S})} \\
\phi_{P} &=\frac{1}{2}\kappa \omega_{p} \int^{t}_{0}\gamma_{p,p}\nn{P(z')}dt' \approx \frac{1}{2} \kappa \omega_{p}\gamma_{p,p} \frac{L}{v_p} \nn{P(z')} \\
\phi_{S} &= \kappa \omega_{p} \int^{t}_{0}\gamma_{s,p}\nn{S(z'+\Delta{v}t')}dt'  \approx \kappa \omega_{p}\gamma_{s,p} \frac{L}{v_p} \nn{S(z')} 
\end{align}
\\
	If the second arm of the interferometer is adjusted for a phase shift of $\pi/2$, the difference in the intensity signal on the two detectors gives the phase, and thus an estimate of the photon number. The total integrated signal energy is $I_{det}=\int\int\int d^3r \epsilon_{0}\epsilon \nn{P(z')}\nn{u_{n}}sin(\phi_{S})$. 
Starting from $N_{s} \hbar \omega_s =  \int^{\infty}_{-\infty} dz' \epsilon_{0} \nn{S(z')} = \int^{\infty}_{-\infty} dz' \epsilon_{0} \nn{S} s(z')$ with $\nn{S}=\frac{N_{s} \hbar \omega_s}{\epsilon_{0} L_{eff,S}}$  (similarly, $\nn{P(z')}=\nn{P} p(z')$ and $\nn{P}=\frac{N_{p} \hbar \omega_p}{\epsilon_{0} L_{eff,P}}$),
%
%
the integral for $I_{det}$ is re-ordered again: 
\begin{align}
\label{eq:int_I}
I_{det} &\approx \int\int\int d^3r \epsilon_{0} \nn{P(z')}  \phi_{S}(z') \epsilon \nn{u_{n}} \\
 & \approx \int dz' \epsilon_{0} \nn{P(z')} \phi_{S}(z') \frac{1}{a} \int_{\Lambda} d^3 r  \epsilon \nn{u_{p}} \\
 & = \int dz' \epsilon_{0} \nn{P(z')} \phi_{S}(z') \\
& \approx \kappa \omega_{p}\gamma_{s,p} \frac{L}{v_p} \frac{N_{s} \hbar \omega_s}{\epsilon_{0} L_{eff,S}} \frac{N_{p} \hbar \omega_p}{ L_{eff,P}} \int dz' p(z') s(z') 
\end{align}
	In the case when P is much narrower than the signal pulse S, we can view it as a quasi delta function $p(z')\approx L_{eff,P} \delta(x-x_0)$. The effective length can be approximated by the spatial width of the pulse, which is $\tau_{s} v_{s}$, the product of the temporal width and the group velocity of S. In this approximation the detected intensity is :
\begin{align}
I_{det} & \approx \kappa \omega_{p}\gamma_{p,s} \frac{L}{v_p} \frac{N_{s} \hbar \omega_s}{\epsilon_{0} \tau_{s} v_{s}} N_{p} \hbar \omega_p \\ 
& =  c n_2 \hbar^2 \omega^{2}_{p} \omega_{s} \gamma_{p,s} \frac{L}{v_p} \frac{N_{s} N_p}{ \tau_{s} v_{s}} 
\end{align}	
	And the phase shift per photon of S is $\frac{I_{det}}{N_p N_s \hbar \omega_p} = c n_{2} \gamma_{p,s} \hbar \omega_{s} \omega_{p} \frac{L}{v_{p}}\frac{1}{v_{s} \tau_{s}}$. In order to make a comparison to our initial plane wave argument, we can identify $\gamma_{p,s}$ as the effective inverse area, $\frac{c}{v_{s} \tau_{s}}$ as the effective pulse bandwidth in the waveguide, and $\frac{L}{v_{p}}$ as the enhanced interaction length (time).   
\bibliographystyle{apsrev}
\bibliography{Kerr_Sub_ARXIV}

\end{document}